# Greatly Enhanced Emission from Spin Defects in Hexagonal Boron Nitride Enabled by a Low-Loss Plasmonic Nano-Cavity

Xiaohui Xu,[†,1,2] Abhishek. B. Solanki,[†,2,3] Demid Sychev,[2,3,4] Xingyu Gao,[4] Samuel Peana,[2,3] Aleksandr S. Baburin,[4,5] Karthik Pagadala,[2,3] Zachariah O. Martin,[2,3] Sarah N. Chowdhury,[2,3] Yong P. Chen,[2,3,6,7,8,9,10] Ilya A. Rodionov,[4,5] Alexander V. Kildishev,[2,3,7] Tongcang Li,[2,3,6,7] Pramey Upadhyaya,[3,7,8] Alexandra Boltasseva,[1,2,3,7,8] Vladimir M. Shalaev[2,3,7,8,*]

[1]School of Materials Engineering, Purdue University, West Lafayette, IN 47907, USA.

[2]Birck Nanotechnology Center, Purdue University, West Lafayette, IN 47907, USA.

[3]Elmore Family School of Electrical and Computer Engineering, Purdue University, West Lafayette, IN 47907, USA.

[4]FMN Laboratory, Bauman Moscow State Technical University, Moscow 105005, Russia.

[5]Dukhov Automatics Research Institute (VNIIA), Moscow 127055, Russia.

[6]Department of Physics and Astronomy, Purdue University, West Lafayette, Indiana 47907, USA.

[7]Purdue Quantum Science and Engineering Institute (PQSEI), Purdue University, West Lafayette, Indiana 47907, USA

[8]The Quantum Science Center (QSC), a National Quantum Information Science Research Center of the U.S. Department of Energy (DOE), Oak Ridge National Laboratory, Oak Ridge, TN 37831, USA

[9]Institute of Physics and Astronomy and Villum Center for Hybrid Quantum Materials and Devices, Aarhus University, 8000 Aarhus-C, Denmark

[10]WPI-AIMR International Research Center for Materials Sciences, Tohoku University, Sendai 980-8577, Japan

[†]These authors contributed equally.

[*]Corresponding author. Email: shalaev@purdue.edu

**Abstract**

Two-dimensional hexagonal boron nitride (hBN) has been known to host a variety of quantum emitters with properties suitable for a broad range of quantum photonic applications. Among them, the negatively charged boron vacancy ($V_B^-$) defect with optically addressable spin states has emerged recently due to its potential use in quantum sensing. Compared to spin defects in bulk crystals, $V_B^-$ preserves its spin coherence properties when placed at nanometer-scale distances from the hBN surface, enabling nanometer-scale quantum sensing. On the other hand, the low quantum efficiency of $V_B^-$ has hindered its use in practical applications. Several studies have reported improving the overall quantum efficiency of $V_B^-$ defects using plasmonic effects; however, the overall enhancements of up to 17 times reported to date are relatively modest. In this study, we explore and demonstrate much higher emission enhancements of $V_B^-$ with ultralow-loss nano-patch antenna (NPA) structures. An overall intensity enhancement of up to 250 times is observed for NPA-coupled $V_B^-$ defects. Since the laser spot exceeds the area of the NPA, where the enhancement occurs, the actual enhancement provided by the NPA is calculated to be ~1685 times, representing a significant increase over the previously reported results. Importantly, the optically detected magnetic resonance (ODMR) contrast is preserved at such exceptionally strong enhancement. Our results not only establish NPA-coupled $V_B^-$ defects as high-resolution magnetic field sensors operating at weak laser powers, but also provide a promising approach to obtaining single $V_B^-$ defects.

Solid-state quantum emitters (QEs) are a central building block for emerging quantum technologies including quantum information processing, quantum communication and quantum sensing[1–4]. In recent years, significant effort has been made with the development of solid-state QEs in two-dimensional van der Waals (vdW) materials like transition metal dichalcogenides (TMDC)[5,6] and hexagonal boron nitride (hBN)[7,8]. The two-dimensional nature of these host materials and their robust chemical properties offer unparalleled advantages for integrating these QEs with plasmonic and photonic structures[9] into hybrid quantum devices. Specifically, hBN has emerged as a promising platform that hosts QEs with remarkable properties such as strain[10] and electric field[11] tunable optical constants, emission ranging from ultraviolet[12] to near infrared[13] wavelengths and spin-selective optical transitions[14]. Consequently, a significant amount of research has been focused on understanding the properties of these defects[15–19], deterministic creation processes[20–24], and their integration with photonic structures[25–28].

Recently, the negatively-charged Boron-vacancy ($V_B^-$) spin defect in hBN[15–19,29–31] has been widely studied for its potential as an atomic scale quantum sensor for magnetic fields, temperature, pressure, and nuclear spins[32–35]. The atomic structure of $V_B^-$ is illustrated in Figure 1a. Previous works have shown that $V_B^-$ has a ground-state splitting energy of $D_{GS}/h$ ~3.5GHz[29,36], and a spin coherence time $T_2$ on the order of 1 μs at ambient conditions[29,37]. The ground state spin population can be initialized and read out optically by virtue of its spin-dependent nonradiative channels and has a resulting optical spin contrast of up to 45%[37] (Figure 1b). It has also been shown that $V_B^-$ defects implanted as shallow as ~ 3 nm from the surface retain their ground state spin coherence properties at ambient conditions[37]. This development, combined with the facile integration of few-layer hBN with other vdW materials, can enable quantum sensing at distances of nanometers from materials of interest.

Despite the great potential of $V_B^-$ defects, their sensitivity to external fields is limited by their poor quantum efficiency[15] and weak photoluminescence (PL) signal. Their low quantum efficiency is also an important reason why single $V_B^-$ defects have not been observed yet. Therefore, improving the overall quantum efficiency by accelerating the photon emission rate, while retaining the spin properties, constitutes a major task in developing practical quantum sensors with $V_B^-$ defects. The photon emission rate of emitters can be enhanced by coupling to plasmonic structures. The intense local electromagnetic field and subwavelength mode confinement enabled by plasmonics increase both the excitation rate and spontaneous emission rate of an emitter, hence producing enhanced fluorescence rates[38–40]. Previous demonstrations have explored this idea by coupling $V_B^-$ defects with nano-patch antennas (NPAs)[41] and gold films[37], with an overall intensity enhancement of up to 17 times. This estimate corresponds to an actual enhancement factor of ~100 in the NPA configuration considering the ratio of laser spot size and NPA area.

It is known that the magnitude of plasmonic enhancement strongly depends on the type and quality of plasmonic materials utilized. By employing suitable plasmonic materials with low optical losses the efficiency of plasmonic cavities can be greatly improved[42], leading to higher Purcell enhancements even for the same cavity configuration. This study demonstrates this effect by coupling $V_B^-$ defects with a resonant nano-plasmonic cavity in the NPA configuration. Unlike the previous works where gold films are employed, we use an epitaxially grown silver film[43,44] with superior optical properties for the NPA fabrication[45,46]. By carefully designing and assembling the structure, we achieve an emission enhancement of $V_B^-$ defects of ~1685 times, which is more than an order of magnitude improvement over previous results (~100)[37,41,47]. The demonstrated Purcell enhancement and preserved ODMR contrast significantly improve the sensitivity of $V_B^-$ for advanced sensing applications. Furthermore, our NPA structure holds great promise for isolating single $V_B^-$ defects, which has not been realized yet and will further extend the application of $V_B^-$ in the field of quantum information.

We focus on hBN flakes with thicknesses of <10 nm containing $V_B^-$ defects induced by helium ion implantation[37]. To assemble the NPA, thin hBN flakes are encapsulated between a single-crystal silver nanocube and a silver film coated with a thin alumina spacer layer. Silver is well known as the best plasmonic material at visible and near infrared frequencies due to the lowest loss at these frequencies among metals[48,49]. We use epitaxially grown silver films of high crystalline quality enabling low intrinsic loss and superior plasmonic enhancement[45,50]. This superior quality is confirmed by atomic force microscopy (AFM) measurements where a film roughness of ~0.3 nm (Figure 1d) is observed, and ellipsometry measurements included in the Supplementary Information. It should be noted that while the exact orientation of the $V_B^-$ optical transition dipole[51] is not *a priori* known, the gap plasmon mode in the NPA configuration is expected to predominantly enhance its vertically-oriented component.

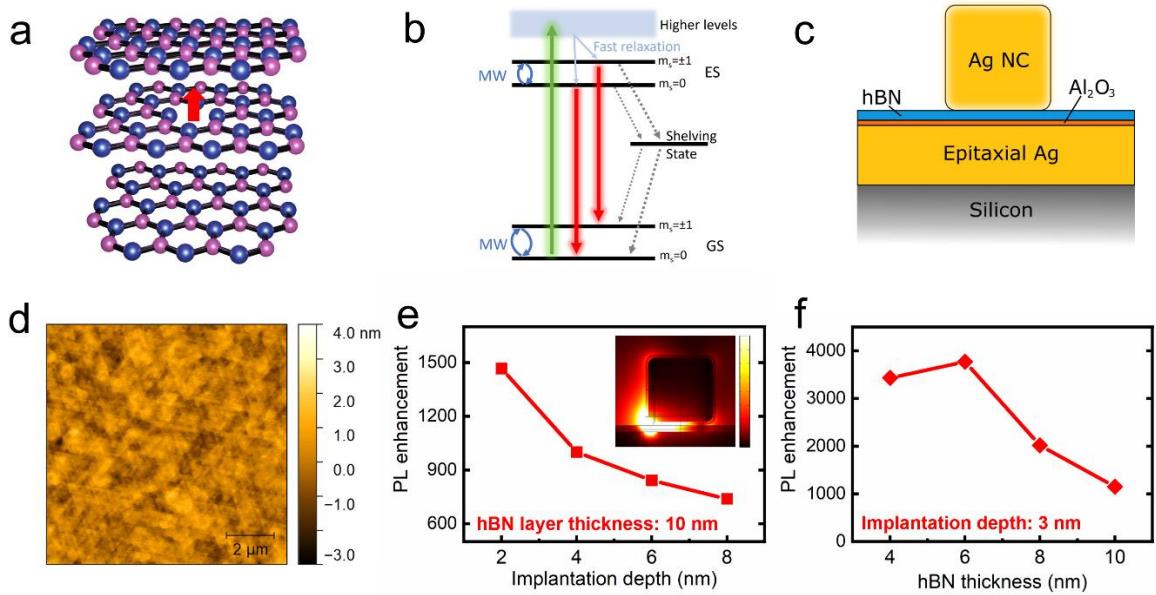

Figure 1. a) Illustration of the atomic structure of $V_B^-$ in hBN. Boron atoms are shown in Purple while Nitrogen atoms shown in Blue. b) A schematic of the electronic spin levels of $V_B^-$. c) The NPA structure adopted in this study, composed of an epitaxial silver film grown on silicon, a 3 nm thick $Al_2O_3$ spacer layer, an hBN layer doped with $V_B^-$ defects and a single-crystal silver nanocube on top. d) AFM scan taken from the epitaxial silver film (65 nm in thickness), revealing a surface roughness of 0.32 nm. e) & f) Simulated PL enhancement factors of an NPA-coupled $V_B^-$ defect at 800 nm as a function of the $V_B^-$ implantation depth (e) and the hBN layer thickness (f). In e), the hBN layer thickness is fixed at 10 nm. Inset: Normalized electric field distribution in a NPA obtained from COMSOL modeling. Color scale range: $0 - 8 \times 10^{32}$ V/m. In f), the implantation depth of $V_B^-$ is kept at 3 nm. All numerical simulations are performed with the finite-element method in the frequency domain (COMSOL Multiphysics), see details in the Supplementary Information.

To achieve the best plasmonic enhancement of $V_B^-$ emission, we consider two parameters in the NPA design using the finite-element method (FEM) in the frequency domain simulations, employing a commercial solver (COMSOL Multiphysics, Wave Optics Module): (i) the implantation depth of $V_B^-$ defects that determines the vertical location of $V_B^-$ in the gap and (ii) the hBN layer thickness which determines the gap size. We calculated the fluorescence enhancement of $V_B^-$ emission (see Supplementary Information) as a function of both parameters and used it as a figure of merit. As shown in Figure 1e, the fluorescence enhancement decreases monotonically with increasing implantation depth for a fixed hBN layer thickness. Intuitively, this can be understood by considering that deeper implantation brings $V_B^-$ defects closer to the underlying silver film, resulting in quenched $V_B^-$ emission coupled with surface plasmon polaritons (SPPs)[52]. On the other hand, shallow implantation positions the $V_B^-$ defects closer to the silver nanocube which efficiently outcouples the radiation into the far field modes. In our experiments, $V_B^-$ defects are implanted at ~3 nm[37] depth from the surface by using the lowest available ion implantation energy (200 eV) on our implanter. As shown in Figure 1f, the fluorescence enhancement increases with decreasing hBN layer thickness and peaks at 6 nm layer thickness. Further reducing the hBN thickness results in decreased fluorescence, likely due to a significant rise in ohmic losses in the metal film[53]. Therefore, hBN flakes that are ~6 nm thick with $V_B^-$ defects implanted at 3 nm from the surface are the most desirable in this study.

We begin with thin hBN flakes that are mechanically exfoliated from high-quality hBN crystals and placed onto $SiO_2$-coated Si substrates. Using the thickness-dependent optical contrast of hBN flakes on $SiO_2$ along with AFM measurements, we identified flat hBN flakes with suitable thicknesses of ~6 nm, with an example shown in Figure 2a (more details in Supplementary Information). The flakes are then implanted with helium ions[37] to induce $V_B^-$ defects. A photoluminescence (PL) map collected from the flake in Figure 2a after ion implantation confirms the formation of a uniform layer of fluorescing $V_B^-$ emitters (Figure 2b). Under 532 nm laser excitation, $V_B^-$ defects show a broad emission spectrum in the range of 700 – 1000 nm and a saturation intensity of 1.14 Mcts/s (Figure 2c), consistent with previous reports[36,37]. The fluorescence lifetime of $V_B^-$ can be extracted from the time-resolved fluorescence decay upon 520 nm pulsed laser excitation, as illustrated in Figure 2c. The fluorescence decay can be fitted using a bi-exponential function with $\tau_1$ = 0.03 ps and $\tau_2$ = 1.18 ns. The slower component, $\tau_2$, is assigned to be the lifetime of $V_B^-$ and agrees well with reported values[41]. The physical origin of $\tau_1$ is not well understood yet but could be from either the instrument itself or ultrafast decays of organic residues on the substrate from the hBN transfer process. Finally, ODMR measurements are performed by delivering microwaves through a copper wire positioned close to the hBN flake (Supplementary Information). An ODMR contrast of ~3% is observed (Figure 2d).

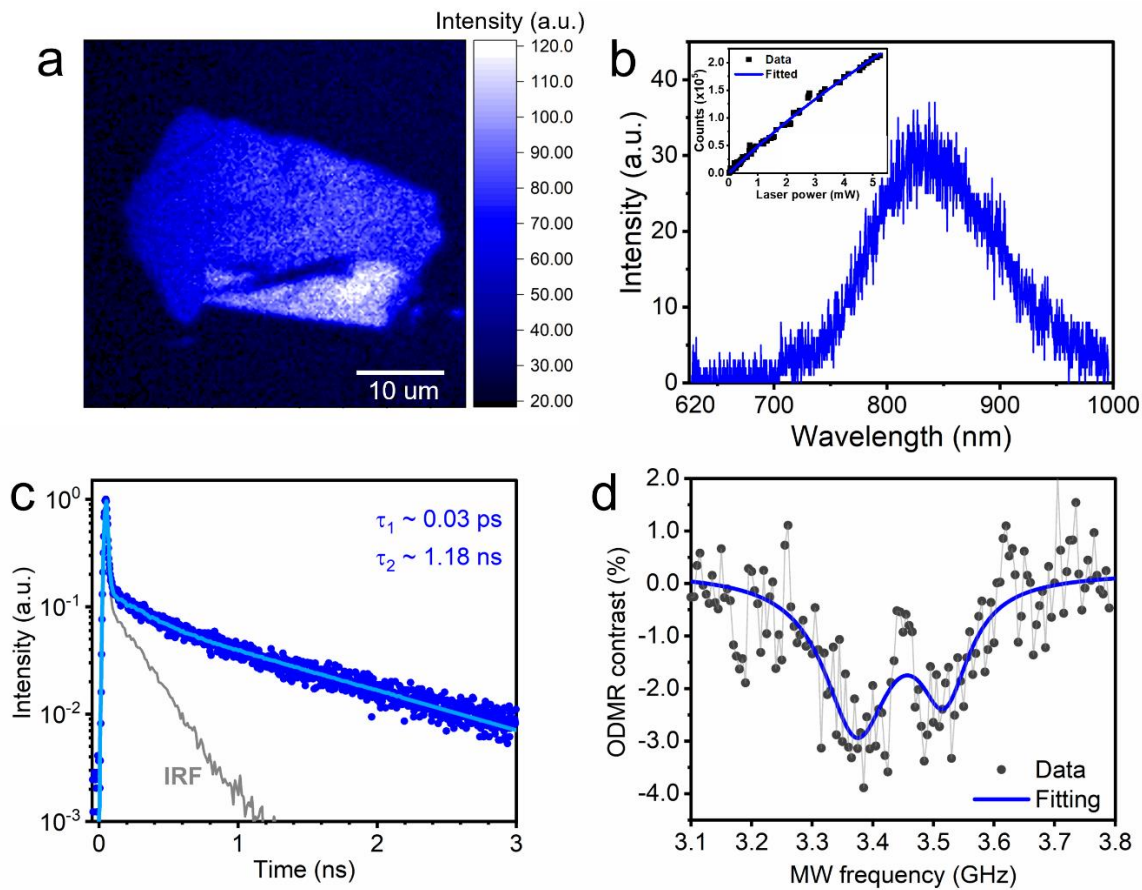

Figure 2. Photophysical properties of $V_B^-$ defects on a SiO$_2$/Si substrate. a) A Photoluminescence (PL) map collected from an hBN flake containing ensembles of $V_B^-$ defects. The map was obtained under 532 nm CW laser excitation at 1 mW power at the objective. The height profile of the same flake could be found in Figure S2 (Supplementary Information), showing a layer thickness of 6.5 nm. b) Background-subtracted emission spectrum of $V_B^-$ defects from the hBN flake shown in a) Inset: background-subtracted fluorescence saturation curve of $V_B^-$ defects with a saturation count of 1.14 Mcps at 22.6 mW. c) Time-resolved fluorescence decay of $V_B^-$ defects (dark blue dots) under the excitation of a 520 nm fs pulsed laser. The data is fitted with a bi-exponential decay function (solid light blue line) convoluted with the instrumental response function (IRF) (solid grey line). Fitted time constants are shown in the graph. d) CW ODMR spectrum of $V_B^-$ (black dots) at 1 mW laser excitation power. Solid blue line is a double-Lorentzian fitting curve, revealing an ODMR contrast of 3%.

After depositing a 3 nm-thick alumina layer on epitaxial silver films (65 nm in thickness), we transfer pre-characterized thin hBN flakes containing $V_B^-$ defects onto silver substrates using polydimethylsiloxane (PDMS) stamps (Supplementary Information). Finally, silver nanocubes with an edge length of ~100 nm are drop-cast to form NPAs on hBN. Figure 3a shows a PL map from an area where a thin hBN flake (6.5 nm in thickness, see Supplementary Information) has randomly distributed NPAs. $V_B^-$ defects in NPAs (referred to as NPA-coupled defects) are then characterized and compared with $V_B^-$ emitters on SiO$_2$/Si substrates (referred to as uncoupled defects). It should be noted that due to the plasmonic enhancement of NPA, a significantly lower laser power is needed for optical characterization. Figure 3b shows the emission spectrum from NPA-coupled defects (circled in red in Figure 3a) along with a spectrum collected from the same hBN flake on SiO$_2$/Si substrates. The emission intensity from NPA-coupled $V_B^-$ defects is significantly higher than that from uncoupled ones, despite the nearly 50-fold smaller laser power used in the former measurement scheme. To better quantify the enhancement factor, saturation curves of NPA-coupled and uncoupled $V_B^-$ emitters are compared in Figure 3c, where an overall intensity enhancement of ~120 times is observed in a broad range of laser powers. This is significantly higher plasmonic enhancement than the previous reports for $V_B^-$ defects[37,41].

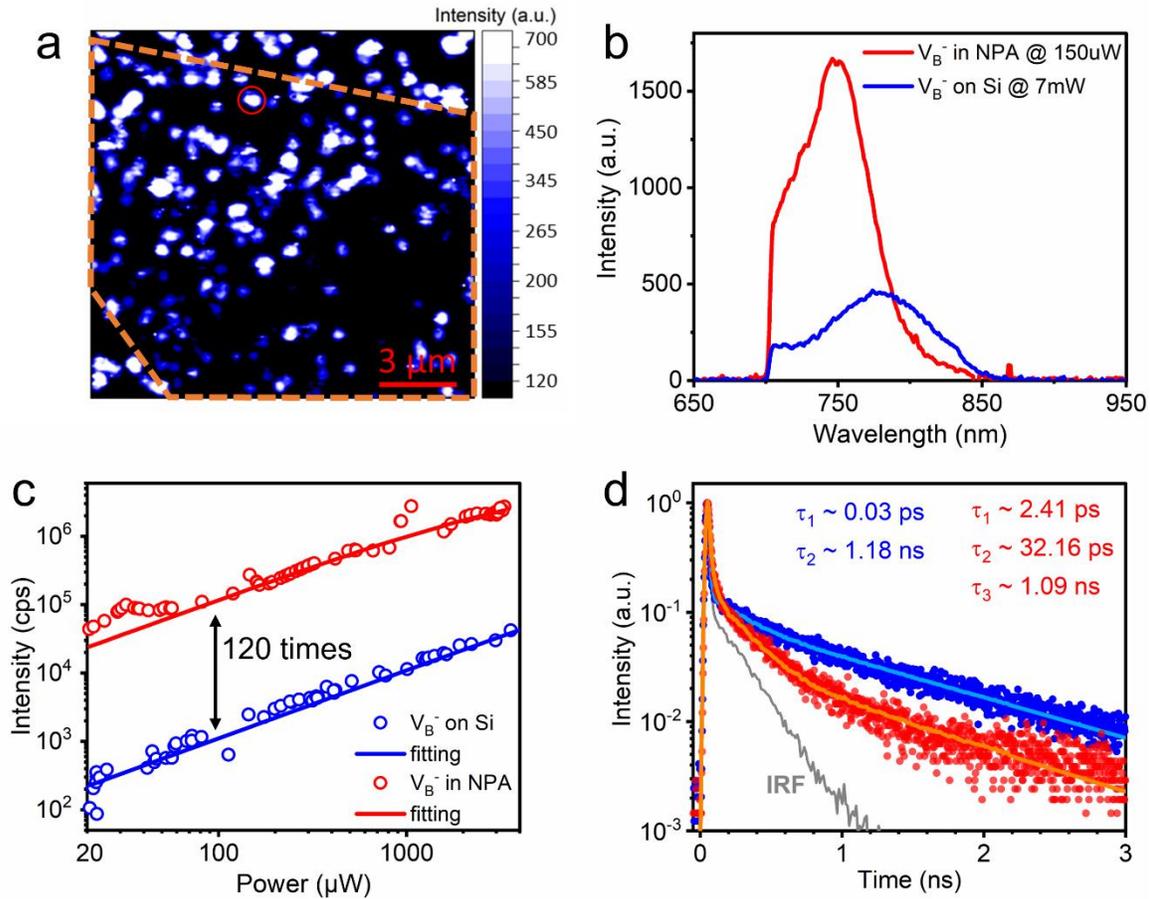

Figure 3. Photophysical properties of NPA-coupled $V_B^-$ emitters. a) PL map taken from part of an hBN flake (enclosed with the orange dashed line) that was implanted with $V_B^-$ defects and sandwiched between randomly formed NPAs. Figure S4 is rescaled to better show the hBN flake under investigation. b) Background-subtracted emission spectra of NPA-coupled $V_B^-$ emitters (solid red line, circled in red in a)) and uncoupled $V_B^-$ emitters from the same flake on a SiO$_2$/Si substrate prior to the transfer (Blue solid line). The corresponding excitation laser powers are indicated in the graph. c) Background-subtracted saturation curves of NPA-coupled (circled in red in a)) and uncoupled $V_B^-$ emitters, shown as red and blue circles, respectively. The solid red and blue lines are fittings. d) Fluorescence decay of NPA-coupled $V_B^-$ emitters (red dots) under pulsed laser excitation, which is fitted using a tri-exponential decay function with time

constants shown in red in the graph. The blue dots and curve are taken from Figure 2c and plotted here for comparison.

The improved brightness of plasmon-enhanced QEs is typically accompanied by lifetime shortening. To gain insight into this, the fluorescence decay curve of $V_B^-$ emitters coupled to the same NPA described above is collected and compared to that from uncoupled $V_B^-$ emitters on SiO$_2$/Si. As shown in Figure 3d, the decay curve cannot be straightforwardly fitted with a bi-exponential decay as in the case of uncoupled $V_B^-$ defects. This difficulty comes from the non-uniform PL enhancements experienced by $V_B^-$ emitters within the excitation laser spot. Note that the silver nanocube forming the NPA has a facet area ($10^4$ nm$^2$) at least an order of magnitude smaller than the pulsed laser spot (spot size > π·260nm$^2$ ≈ 2.1×10$^5$ nm$^2$ at a wavelength of 520 nm). Hence, both NPA-coupled and uncoupled $V_B^-$ emitters are excited during the time-resolved measurements. This observation is supported by the fact that the decay curve is well fitted using a tri-exponential function with $\tau_1$ = 2.41 ps, $\tau_2$ = 33.16 ps and $\tau_3$ = 1.09 ns. Again, the fastest component $\tau_1$ is unrelated to the $V_B^-$ decay dynamics. $\tau_2$ and $\tau_3$ can be attributed to lifetimes of strongly enhanced (NPA-coupled) and weakly enhanced (uncoupled) $V_B^-$ emitters, respectively. Therefore, the lifetime shortening of NPA-enhanced $V_B^-$ is 36 times. As expected, this is smaller than the PL enhancement factor due to the high intrinsic nonradiative decay rates of $V_B^-$ [38]. The uncoupled $V_B^-$ emitters on the silver film show a slight lifetime shortening (1.1 times). As discussed later, this enhancement comes from the silver film's localized surface plasmon modes (LSPs).

For practical quantum sensing applications, it is critical to ensure that the spin contrast and PL intensity are significant. Due to the high reflectivity of the silver substrate and silver nanocubes, microwave delivery by a copper wire on the sample surface is a nonviable solution. We circumvent this issue by using lithographically patterned waveguide-mediated microwave delivery as reported in previous studies[33,34,37]. A coplanar waveguide structure offers more efficient control of the defect spins with the 'in-plane' orientation of the microwave fields driving the 'out-of-plane' spins.

Figure 4a shows an optical image of the silver waveguide (see the design in Figure S5, Supplementary Information). To achieve the best ODMR results, hBN flakes containing $V_B^-$ defects are transferred onto the central neck region of the waveguide with a width of 30 μm. Notably, the quality of epitaxial silver films is mostly preserved after all fabrication steps, as confirmed by AFM (Figure 4a, bottom). We attribute this to the protection of a photoresist layer throughout the fabrication process and the lack of any high-temperature treatments. The preserved silver film quality also ensures plasmonic performance comparable to that of the previous section. We transfer an implanted hBN flake with a thickness of 6.5 nm to the waveguide. As shown by the AFM image in Figure 4b, the hBN flake is successfully placed onto the waveguide, although the transfer process introduces additional folds/wrinkles to the flake. Nevertheless, flat regions on the flake are suitable for NPA fabrication and are used in the following study. Figure 4c is a PL map of the flake showing well-isolated NPAs after drop-casting silver nanocubes. Specifically, the NPA circled in red is studied systematically, with its spectrum and saturation curve plotted in red in Figure 4d and 4e, respectively, along with the data collected from the same flake on a SiO$_2$/Si substrate (blue curves). Again, we observe a significant overall emission intensity enhancement from NPA-coupled $V_B^-$ emitters, around 250 times at low laser powers (<300 uW). The enhancement factor slightly decreases at higher laser powers as NPA-coupled $V_B^-$ emitters become saturated earlier than the uncoupled ones. The fluorescence decay from the area containing the NPA circled in red was also measured. The details of the fitting procedure and results are presented in the Supplementary Information. Finally, ODMR measurements are performed by exciting the hBN sample with microwaves through the silver waveguide. Figure 4f shows an ODMR contrast of ~6% is observed from NPA-coupled $V_B^-$ emitters.

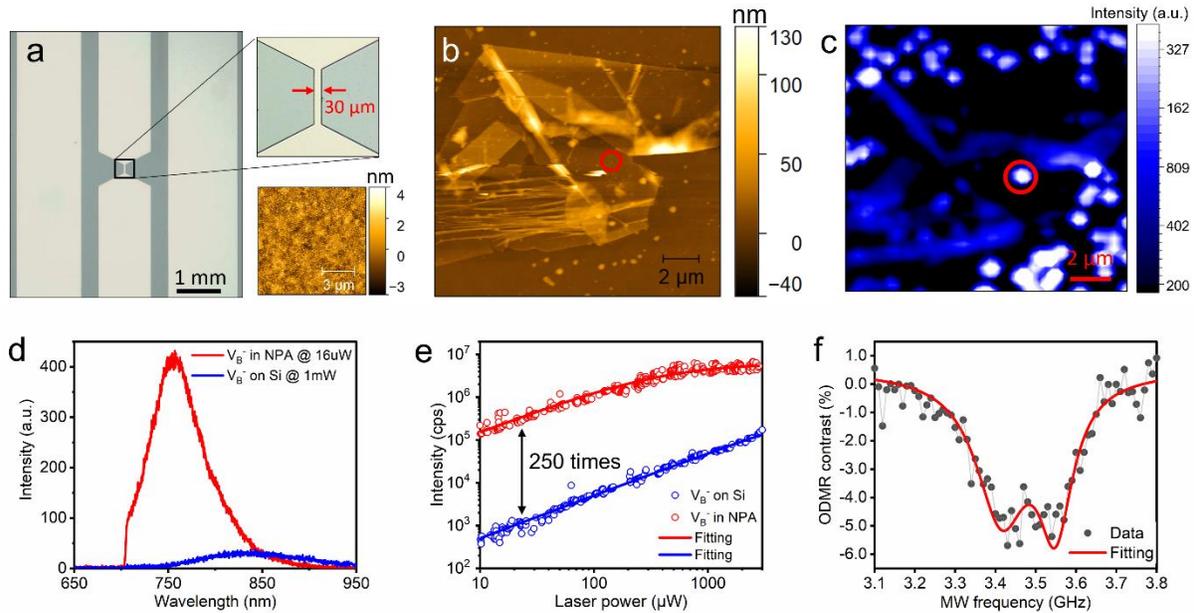

Figure 4. a) Left: optical image of the microwave waveguide fabricated on a 300 nm-thick silver film grown epitaxially on silicon. The bright and dark regions represent silver and silicon, respectively. Top right: A zoom-in optical view of the central neck region of the waveguide, showing a neck width of 30 μm. Bottom right: AFM scan from the neck region of the waveguide. A surface roughness of 0.91 nm is obtained. b) AFM scan of an hBN flake after it is transferred onto the neck region of the microwave waveguide. This is the same flake shown in Figure 2a. c) PL map of the area containing the hBN flake after drop-casting silver nanocubes, collected along the same orientation as in b). Due to stronger fluorescence from folds/wrinkles on hBN, the flake location can be identified by matching the fluorescence pattern with the AFM map in b). The NPA circled in red is also circled in b) to help understand the matching of two maps. d) Background-subtracted emission spectra of NPA-coupled $V_B^-$ emitters (solid red line, circled in red in a)) and uncoupled $V_B^-$ emitters from the same flake on a SiO$_2$/Si substrate prior to the transfer (Solid blue line). The corresponding excitation laser powers are indicated in the graph. e) Background-subtracted saturation curves of NPA-coupled and uncoupled $V_B^-$ emitters, shown as red and blue circles, respectively. The solid

red and blue lines are fittings. An emission enhancement of ~250 times is observed at laser powers < 300 µW. f) CW ODMR spectrum of $V_B^-$ (black dots) at a 40 mW microwave driving power and 30 µW laser excitation power. The solid red line is a double-Lorentzian fitting curve that yields an ODMR contrast of 6%.

While the intensity enhancement obtained with the waveguide structure is higher than that from the NPA on a planar silver film discussed earlier, the underlying physics is fundamentally the same. The difference in enhancement factor can be attributed to the local variation in the optical properties of the NPA structure such as the roughness of silver, roughness of hBN, quality of silver nanocubes, etc. This is also reflected in the variation of enhancement factors that we observed experimentally from several NPAs located on the same hBN flake (details discussed in the Supplementary Information). Nevertheless, the significant plasmonic enhancement combined with the ODMR contrast from the NPA-coupled defects in the waveguide again confirms the high quality of epitaxial silver films after lithography processes and presents a significant advancement for sensing applications.

In addition to optical characterization of NPA-coupled $V_B^-$ emitters, we also examined $V_B^-$ emitters that are only coupled with the silver film. This approach helps to isolate contributions from the silver film and the NPA structure to the observed fluorescence enhancement. To this end, we measured a spot (circled in orange in Figure S8a) on hBN close to the NPA characterized above. The spectrum of the silver film-coupled $V_B^-$ emitters is plotted in Figure S8b, with an emission intensity slightly higher than that of uncoupled $V_B^-$ defects on sSiO$_2$/Si under the same pump laser power. A 6-fold intensity enhancement by the silver film at low laser powers is extracted by comparing the collected saturation curves (Figure S8c). The main mechanism of plasmon-enhanced emission by a silver film is the excitation of localized surface plasmon (LSP) modes on the film surface. Ideally, plasmonic enhancement is not expected from smooth, crystalline metal films which only support propagating surface plasmon modes that weakly couple to photons due to momentum mismatch. In practice, however, even high-quality metal films show roughness at the nanoscale. Such local features could potentially relax the momentum mismatch condition and lead to LSP excitation. This is the suggested mechanism for the plasmon-enhanced $V_B^-$ emission in ref[37], where a 17-fold intensity enhancement by a gold film is reported. The relatively lower enhancement by our silver film could be attributed to the metal film's higher quality and lower roughness. It also indicates that the gap plasmon modes in the NPA play a dominant role in enhancing $V_B^-$ emission in this work, rather than the LSP modes in the silver film.

From the discussion above, we are better positioned to quantify the actual enhancement factor by the NPA structure. Again, only $V_B^-$ defects under the silver nanocube are enhanced by the gap plasmon mode of NPA, and the rest of the defects within the pump laser spot are enhanced just by the silver film. Hence, the actual enhancement factor by NPA can be estimated based on the following expression:

$$\frac{P_{NPA} \cdot A_{cube} + P_{Ag} \cdot (A_{laser} - A_{cube})}{A_{laser}} = P_{ave}$$

where $P_{Ag} = 6$ and $P_{ave} = 250$ refer to the enhancement factor by the silver film and the overall enhancement factor measured experimentally from the laser spot area containing an NPA, respectively; $A_{cube}$ and $A_{laser}$ are the silver nanocube facet area ($10^4$ nm$^2$) and laser spot area. For a 532 nm pump laser under optimum alignments, its spot size can be regarded as diffraction-limited and is estimated to be 296 nm according to the Abbe limit (considering a numerical aperture of 0.9). Combining the values above, the actual enhancement factor by NPA denoted as $P_{NPA}$ is calculated to equal 1685. While the calculation above does not include factors like the exact optical dipole orientation of $V_B^-$, the spatial heterogeneity of $V_B^-$ emitters in hBN, the effect of laser beam shape, etc., it clearly demonstrates an intensity enhancement of $V_B^-$ emission by more than three orders of magnitude in NPA. The achieved enhancement significantly improves the overall quantum efficiency of $V_B^-$ by boosting its radiative decay rate, hence providing a promising way to obtain single $V_B^-$ emitters with observable brightness. To explore this, future works can combine the silver-based NPAs with hBN flakes implanted with smaller ion doses, such that $V_B^-$ defects are appropriately isolated from each other.

We observe slightly higher ODMR contrasts from NPA-coupled emitters (6%) and silver film-coupled emitters (11%, Figure S8d, Supplementary Information) as compared to $V_B^-$ emitters on SiO$_2$/Si (~3%). However, the higher contrasts cannot be attributed to plasmonic enhancement in this case, as the effective microwave field is substantially higher for the waveguide structure. Further investigation is needed to understand the effect of plasmonic enhancement on the spin contrast.

The most important figure of merit in practical sensing applications is the sensitivity to external stimuli. For example, in quantum sensing of external magnetic fields, the sensitivity is defined as the smallest magnetic field that can be measured in one second[4]. In the ODMR-based sensing scheme[4], the sensitivity ($S$) is closely related to the number of photons detected ($N$), the ODMR contrast ($C$) and the linewidth of the ODMR signal ($\Delta v$) as $S \sim \frac{\Delta v}{\sqrt{N} C}$. A higher microwave/laser power can lead to a higher contrast/PL strength at the cost of deteriorating the ODMR linewidth and sample heating. In our experiment, we demonstrate a PL enhancement of ~ 1685 times without significant deteriorations of the ODMR linewidth ($\Delta v$ ~100MHz) and spin contrast ($C$ ~6%) for NPA-coupled $V_B^-$ defects. This leads to an improvement in $S$ of about $\sqrt{1685}$~41 times compared to uncoupled $V_B^-$ defects. For the CW-ODMR scheme, the absolute value of S is calculated to be ~$138\frac{\mu T}{\sqrt{Hz}}$ for NPA-coupled $V_B^-$ at a laser power of 30 µW and microwave power of 40 mW with known parameters[4,54]. This is comparable to the sensitivity of $V_B^-$ reported by Gottscholl *et al*[32]. Gao *et al*[37] has reported higher sensitivities of $V_B^-$ after optimizing the laser and microwave powers. However, note that they were achieved at a much higher laser power (> 1 mW) and microwave power (0.25 W), as well as a larger hBN flake thickness (35 nm). Compared to the reports above, our configuration processes the following unique advantages: 1) a much smaller laser power required for achieving effective sensing, which avoids the detrimental effects of high driving powers; 2) a much higher spatial resolution as defined by the silver nanocube dimension (~100 nm) and an extremely small separation distance between the probing spin and material of interest (3 nm). In the case of sensing magnetic excitations[55], this sample-probe distance not only improves the strength of stray magnetic fields but also determines the wavelength of the excitations that couple with the spin transition. In our configuration, ultrathin hBN flakes can measure stray magnetic fields emanating from magnetic materials at a close distance of ~3 nm, enabling advanced probes of previously unexplored phenomena. In a future study, the sensitivity of NPA-coupled $V_B^-$ emitters can be further improved by optimizing the laser/microwave powers and the implantation doses.

**Conclusion**

In this work, we have demonstrated fluorescence enhancement of ensembles of $V_B^-$ defects in thin hBN flake by coupling them to silver-based nanopatch antennas (NPAs). An overall PL intensity enhancement of up to 250 times is observed for NPA-coupled $V_B^-$ defects. This corresponds to a 1685-fold actual PL enhancement by the NPA when accounting for the laser spot size and the NPA cross-section area. The demonstrated PL enhancement is more than an order of magnitude higher than previous reports on plasmon-enhanced $V_B^-$ defects. An important factor in this enhancement is the use of epitaxial silver films in our NPA fabrication, enabling ultralow optical losses and hence higher Purcell factors. Time-resolved fluorescence measurements revealed a 36-time lifetime shortening for NPA-coupled $V_B^-$ emitters. Furthermore, $V_B^-$ defects retain an ODMR contrast of 6% after being coupled to NPAs. The preserved ODMR contrast and the significant PL enhancement make NPA-coupled $V_B^-$ effective quantum sensors. This is confirmed by our sensitivity evaluation of NPA-coupled $V_B^-$ emitters, where a magnetic field sensitivity of $138\frac{\mu T}{\sqrt{Hz}}$ is calculated at a laser power of 30 µW. The small thickness of hBN flakes (<10 nm) involved in the NPA configuration also offers unique advantages to probing weak fields originating from regions very close to the sample of interest. Aside from advanced sensing applications, the significant PL enhancement achieved in our NPA structure provides a promising way to obtain single $V_B^-$ spin defects, an important step to exploit the potential of $V_B^-$ in quantum information science and technologies.

## Acknowledgment

This work is supported by the U.S. Department of Energy (DOE), Office of Science through the Quantum Science Center (QSC), DE-AC05-00OR22725, National Science Foundation Award 2015025-ECCS and Air Force Office of Scientific Research grant "Hybrid, Room-Temperature, Quantum On-Chip Photonic Systems: Integrating Quantum Emitters with Nanoplasmonics".